\documentclass[a4paper, amsfonts, amssymb, amsmath, reprint, showkeys, nofootinbib, twoside,superscriptaddress]{revtex4-2}
\usepackage[english]{babel}
\usepackage[utf8]{inputenc} 
\usepackage[colorinlistoftodos, color=green!40, prependcaption]{todonotes}
\usepackage{esvect}
\usepackage{parskip}
\usepackage{xcolor}
\usepackage{soul}
\sethlcolor{yellow}
\usepackage{ulem}

\usepackage{float}
\usepackage{upgreek}

\usepackage{fancyhdr}

\usepackage[pdftex, pdftitle={Article}, pdfauthor={Author}]{hyperref} 

\usepackage{totalcount}

\begin{document}

\title{Nitrogen-vacancy magnetometry of CrSBr by diamond membrane transfer}

\author{T.~S.~Ghiasi}\email[Correspondence: ]{t.s.ghiasi@tudelft.nl} \affiliation{Kavli Institute of Nanoscience, Delft University of Technology, 2628 CJ Delft, The
Netherlands}
\author{M.~Borst}  \affiliation{Kavli Institute of Nanoscience,
Delft University of Technology, 2628 CJ Delft, The
Netherlands}
\author{S.~Kurdi}\affiliation{Kavli Institute of Nanoscience,
Delft University of Technology, 2628 CJ Delft, The
Netherlands} 
\author{B.~G.~Simon}\affiliation{Kavli Institute of Nanoscience,
Delft University of Technology, 2628 CJ Delft, The
Netherlands}
\author{I.~Bertelli}\affiliation{Kavli Institute of Nanoscience,
Delft University of Technology, 2628 CJ Delft, The
Netherlands}
\author{C.~Boix-Constant}\affiliation{Institute of Molecular Science, University of Valencia, Catedrático José Beltrán 2, Paterna 46980, Spain}
\author{S.~Ma\~{n}as-Valero}\affiliation{Kavli Institute of Nanoscience, Delft University of Technology, 2628 CJ Delft, The Netherlands}\affiliation{Institute of Molecular Science, University of Valencia, Catedrático José Beltrán 2, Paterna 46980, Spain}

\author{H.~S.~J. van der Zant} \affiliation{Kavli Institute of Nanoscience,
Delft University of Technology, 2628 CJ Delft, The
Netherlands}
\author{T.~van der Sar}
\email{T.vanderSar@tudelft.nl}\affiliation{Kavli Institute of Nanoscience,
Delft University of Technology, 2628 CJ Delft, The
Netherlands} 

\date{\today} 

\begin{abstract}
Magnetic imaging using nitrogen-vacancy (NV) spins in diamonds is a powerful technique for acquiring quantitative information about sub-micron scale magnetic order. A major challenge for its application in the research on two-dimensional~(2D) magnets is the positioning of the NV centers at a well-defined, nanoscale distance to the target material required for detecting the small magnetic fields generated by magnetic monolayers. Here, we develop a diamond `dry-transfer' technique akin to the state-of-the-art 2D-materials assembly methods and use it to place a diamond micro-membrane in direct contact with the 2D interlayer antiferromagnet CrSBr. We harness the resulting NV-sample proximity to spatially resolve the magnetic stray fields generated by the CrSBr, present only where the CrSBr thickness changes by an odd number of layers. From the magnetic stray field of a single uncompensated ferromagnetic layer in the CrSBr, we extract a monolayer magnetization of \textcolor{black}{$M_\mathrm{CSB}=0.46(2)$~T}, without the need for exfoliation of monolayer crystals or applying large external magnetic fields. The ability to deterministically place NV-ensemble sensors into contact with target materials and detect ferromagnetic monolayer magnetizations paves the way for quantitative analysis of a wide range of 2D magnets assembled on arbitrary target substrates.

\end{abstract}

\maketitle
\begin{center}
{\large {\textbf{Introduction}}}
\end{center}
\begin{figure*}[ht!]
    \centering
    \includegraphics[width=\textwidth]{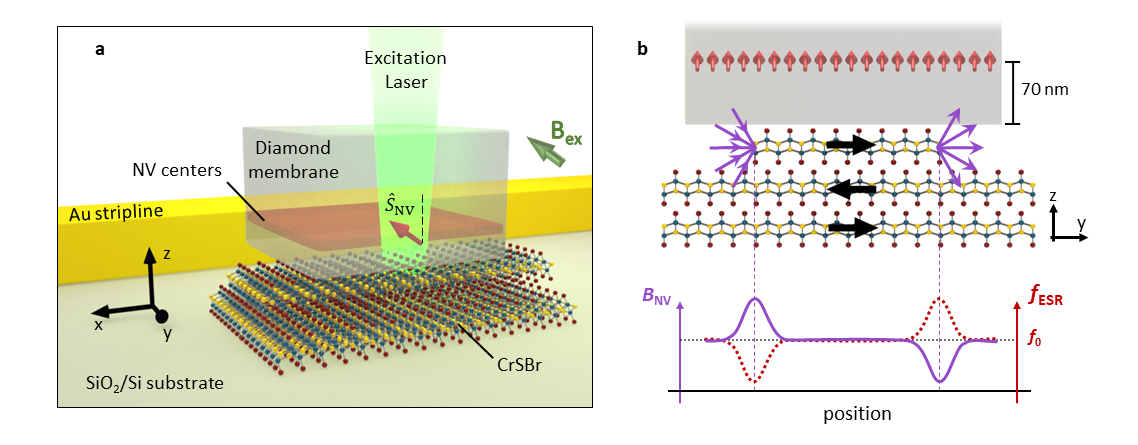}
    \caption{NV magnetometry of the interlayer antiferromagnet CrSBr. (a) Schematics of the sample. A diamond membrane with a shallowly implanted layer of NV centers (at the depth of \textcolor{black}{about} 70~nm, shown with the layer in red) is placed onto the CrSBr crystal. The NV centers consist of a substitutional nitrogen atom adjacent to a vacancy in the diamond lattice. They are oriented along the 111 directions in the diamond crystal, at $\alpha=$~54.7~degrees with respect to the sample normal. We use the subset of NVs oriented along $\hat{S}_\mathrm{NV}=\textrm{sin}\alpha~\hat{x}+\textrm{cos}\alpha~\hat{z}$ for magnetometry. A small external magnetic field $B_\mathrm{ex}=5.6$~mT along $\hat{S}_\mathrm{NV}$ enables selective driving of the $m_\mathrm{s}=0\leftrightarrow-1$ electron spin resonance (ESR) transition of the target NV subset using microwaves applied via the stripline. A green laser (520~nm) excites the spin-dependent NV photoluminescence (PL). (b) Illustration of the modulation of the NV ESR frequency by the stray fields (purple arrows) generated by the uncompensated CrSBr magnetization at the CrSBr edges. The CrSBr magnetic stacking order is indicated by black arrows. The plot shows the expected spatial dependence of the CrSBr stray fields and the corresponding change of the NV ESR frequency ($f_\mathrm{ESR}=f_0 - \gamma dB_\mathrm{NV}$) sensed by the selected NV family, where $f_0 = D-\gamma B_\mathrm{ex}$ with $D$ = 2.87 GHz the NV zero-field splitting. } 
        \label{fig1:my_label}
\end{figure*}

\begin{figure*}[ht!]
    \centering
    \includegraphics[width=\textwidth]{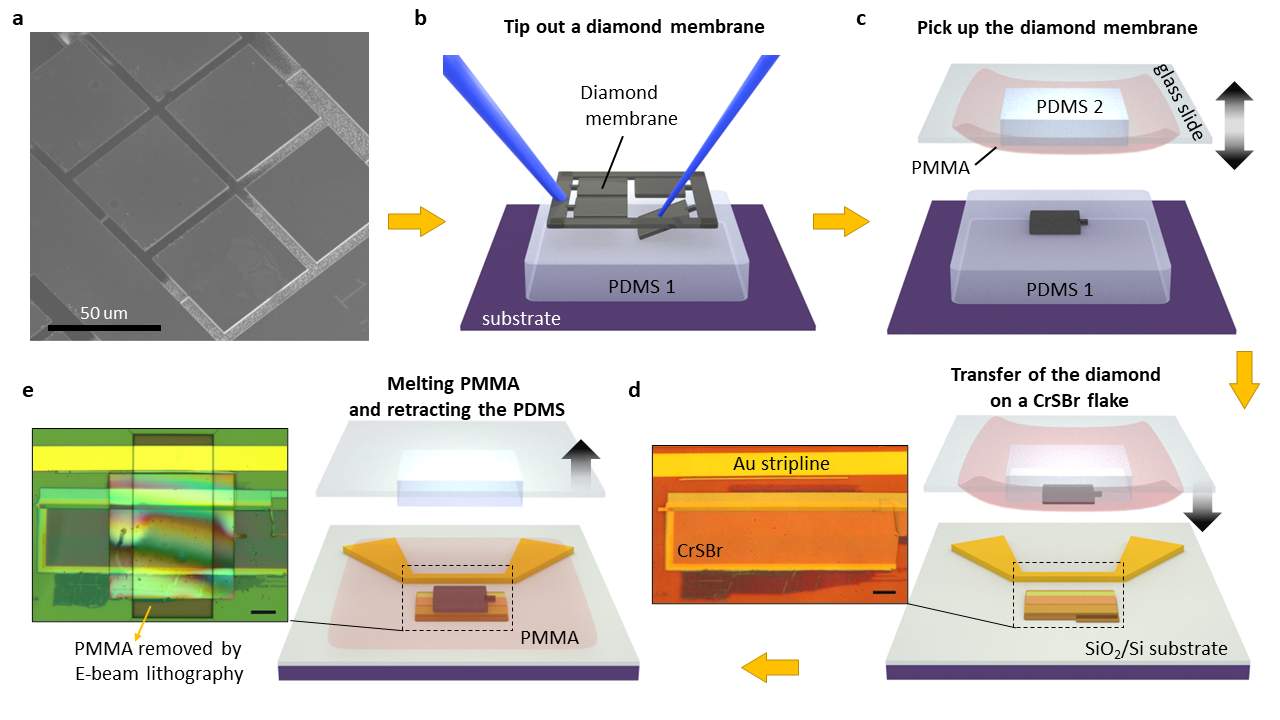}
    \caption{Fabrication of the diamond membrane and transfer on 2D magnets. (a) An SEM image of the diamond membrane etched into the 50$\times$50$\times$5~$\upmu$m$^3$ squares by O$_2$ plasma. (b) Detaching a single square of the diamond-membrane by a metallic tip (shown in blue) that is controlled by micro-manipulators. After the detachment, the membrane drops on a polydimethylsiloxane (PDMS) layer that is held on a substrate (here called PDMS 1). (c) A PMMA-PDMS stamp (prepared separately by suspension of PMMA membrane, described in the Methods section) is vertically brought in contact with the diamond membrane by a micro-manipulator. Adhesion of the diamond to the PMMA allows for the pick-up of the diamond membrane by retracting the PMMD-PDMS stamp. (d) Transfer of the diamond on top of a CrSBr flake that is located next to the Au stripline on a SiO$_2$-Si substrate. The optical micrograph shows the region defined by the black dashed line (the scale bar is 10 $\upmu$m).  (e) By melting the PMMA at 180~$^\circ$C the PMMA and diamond membranes are released on top of the SiO$_2$ substrate, leaving the diamond on the target CrSBr flake within a few micron precision. As shown in the optical micrograph, the PMMA membrane is removed by e-beam lithography at the region of interest.   }
    \label{fig2:my_label}
\end{figure*}

\begin{figure*}
    \centering
    \includegraphics[width=1\textwidth]{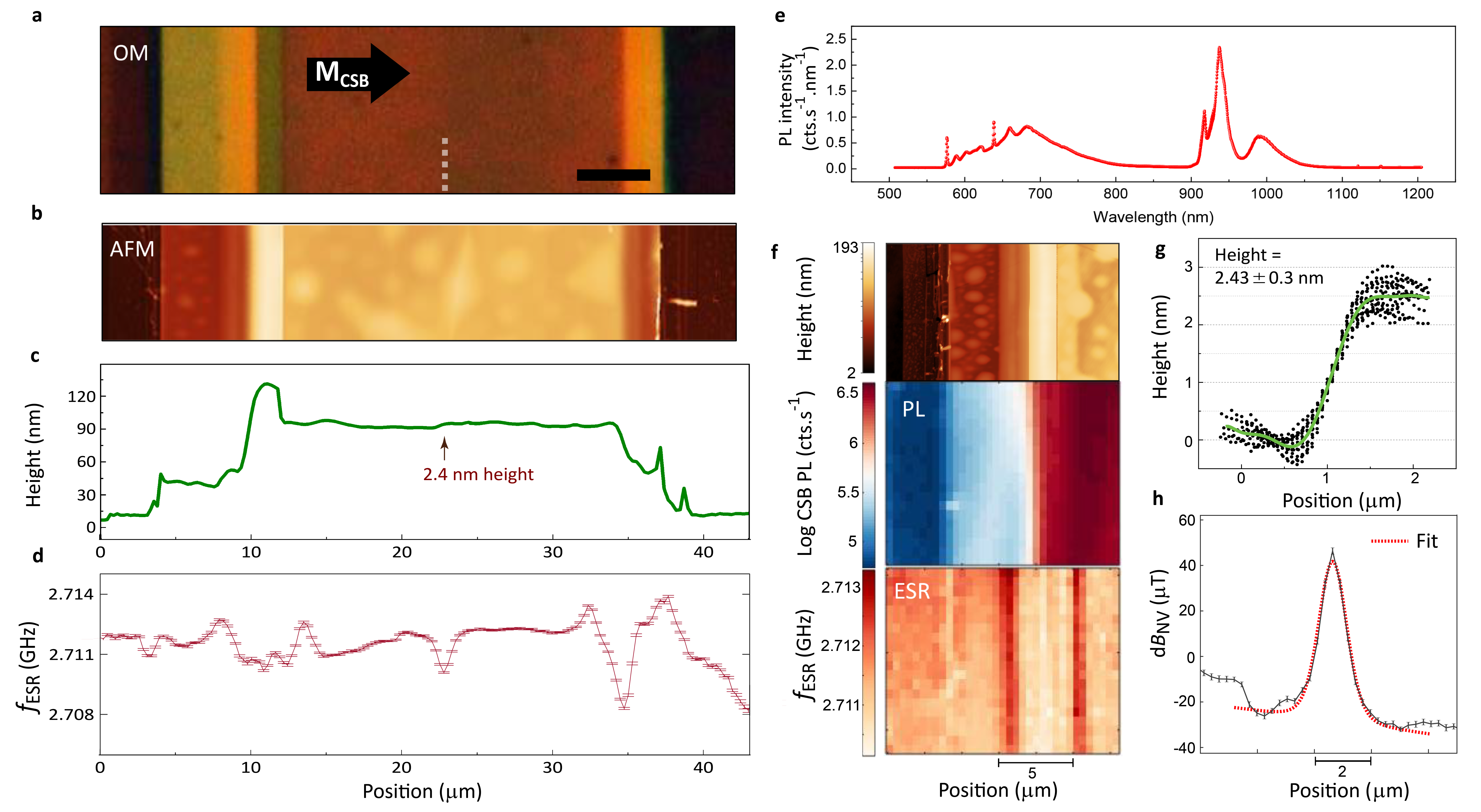}
    \caption{(a) Optical micrograph (OM) of the studied region of the CrSBr flake (scale bar is $5\upmu$m). The black arrow shows the direction of the magnetization easy axis of the CrSBr flake. The white dashed line highlights the edge of the \textcolor{black}{2.4 nm} CrSBr layer. (b) Atomic-force microscope (AFM) image of the CrSBr flake. (c) Height profile of the CrSBr flake, measured along its width. (d) Electron spin resonance frequency measured at $T$= 10 K along the width of the CrSBr flake. \textcolor{black}{The ESR signal is measured along the white dashed lines shown in Fig.~S1 and S3 in the SI.} (e) Photoluminescence (PL) measurements of the NV-diamond and CrSBr, performed at 10~K, with excitation wavelength and power of 520~nm and 10$~\upmu$W, respectively. \textcolor{black}{Due to the high density of the NV centers, the PL spectrum between 600-800 nm is dominated by the PL signal from the NV centers.}  
    (f) Spatial map of the PL and ESR contrast, measured at $T$=~70~K, corresponding to the CrSBr region shown in the AFM image in this panel (the AFM image is presented in an adaptive nonlinear color range to highlight the small steps). (g) AFM height profile \textcolor{black}{of the 2.4 nm} step in the CrSBr flake. \textcolor{black}{The data is collected from 15 individual height profiles collected along the edge of the CrSBr step. The height of 2.43$\pm0.3\,$nm is extracted by subtraction of the mean height values at the two height levels.} (h) Estimated stray field at the \textcolor{black}{2.4 nm} CrSBr edge (solid \textcolor{black}{black} line) and fit (dotted red line, see SI, section S3).  }
    \label{fig3:my_label}
\end{figure*}

The recent emergence of 2D magnetic materials with their potential applications in spin-logic circuitry and memory devices has triggered experimental research to find methods for detection and control of their magnetic ordering~\cite{gibertini2019magnetic}. This has been a challenge for the past decade because of the low magnetic moment of these magnets at the 2D limit, compared with their bulk counterpart. There have been many techniques introduced so far, commonly used for the detection of magnetic behavior and quantitative study of the magnetism in the 2D magnets, such as magneto-transport measurements~\cite{jiang2018electric, song2018giant, wang2018very} and electron tunneling~\cite{klein2018probing, wang2018tunneling}, magnetic circular dichroism and magneto-optical Kerr effect measurements~\cite{huang2017layer, gong2017discovery,fei2018two}, optical second harmonic generation~\cite{sun2019giant, chu2020linear, lee2021magnetic}, magnetic force microscopy~\cite{yi2016competing, niu2019coexistence, rizzo2022visualizing}, scanning superconducting quantum interference device (SQUID) microscopy~\cite{uri2020nanoscale,marchiori2022nanoscale}, angular electron spin resonance~\cite{moro2022revealing} and nitrogen-vacancy~(NV) magnetometry~\cite{maletinsky2012robust, simpson2016magneto, casola2018probing, thiel2019probing, broadway2020imaging,sun2021magnetic,fabre2021characterization, song2021direct,laraoui2022opportunities,robertson2022imaging}. 
Among them, NV-magnetometry provides high sensitivity and nano-precision in the detection of magnetic ordering at a large temperature range from \textcolor{black}{0.35 to 600 K \cite{scheidegger2022scanning, toyli2012measurement}}. By this technique, we can detect weak static and dynamic magnetic stray fields that provide quantitative information about the magnetization of a 2D magnet down to the monolayer limit, providing insights into the magnetic domains and localized magnetic defects~\cite{laraoui2022opportunities}. 

A central challenge for achieving high spatial resolution and extracting quantitative results on material magnetization using NV magnetometry is to achieve a nanoscale and well-defined distance between the sample and the NV centers~\cite{casola2018probing}. A powerful method is to embed a single NV spin into a diamond scanning probe, which enables imaging with 50~nm spatial resolution and quantitative determination of monolayer magnetizations~\cite{thiel2019probing, sun2021magnetic}. A second approach is to deposit a 2D magnet directly onto a diamond chip hosting a high density of near-surface NV spins~\cite{broadway2020imaging,chen2022revealing,mclaughlin2022quantum}\textcolor{black}{\cite{garsi2021non, yan2022quantum, chen2023above}}. This approach benefits from a strong signal due to a large number of NV spins. Furthermore, it enables large-area magnetic imaging with 100~$\upmu$m-scale field-of-views at diffraction-limited resolution~\cite{scholten2021widefield}, but requires the ability to fabricate the sample onto the diamond.

For the NV-magnetometry of 2D materials, here we develop a method based on a deterministic placement of a diamond membrane onto the target magnetic flake on a substrate with micrometer lateral precision. We demonstrate a pick-and-place diamond transfer procedure similar to the `dry-transfer' technique commonly used in assembling stacks of 2D materials~\cite{zomer2014fast}. Using this technique we achieve \textcolor{black}{close proximity} between the diamond and the sample \textcolor{black}{that is crucial for detecting the weak magnetic fields of atomically thin magnetic layers with high spatial resolution. By using micron-sized diamond membranes as done in Ref.\cite{schlussel2018wide}, we strongly increase the probability to achieve nanoscale stand-off distance between diamond and sample with respect to using millimeter-sized diamonds~\cite{bertelli2020magnetic}. Such large diamonds generally lead to micron-scale diamond-sample distances because of the $\sim$3 orders of magnitude larger probability to capture spurious particles (e.g. dust) between diamond and sample. Moreover, our transfer technique positions the diamond at a target location with a micrometer lateral precision,
while being compatible with 2D materials assembly techniques. This high degree of control also precludes an incorrect orientation of the diamond, as well as the need to further push/drag the diamond to a target location using micromanipulators, which can damage sensitive samples such as 2D materials. Furthermore, thinning the diamond membrane to a few microns enhances the collection of the NV photoluminescence because of reduced optical aberrations and yields a flexible membrane that can conform more readily to a target surface, which facilitates pickup with the PDMS/PMMA stamp.} An additional benefit of our approach is that it enables placing a diamond sensor onto commonly used substrates such as SiO$_2$/Si that facilitates optical detection and thickness determination of 2D materials. Moreover, these substrates and materials can be equipped with electronic circuitry (e.g. for the microwave resonator) to study and control the various spins in the system. 

We use the diamond transfer technique to characterize the magnetization of a 2D interlayer antiferromagnet, Chromium Sulfur Bromide (CrSBr) (Fig.~1a). The diamond membrane contains a layer of NV centers implanted at \textcolor{black}{about} 70~nm below the surface (see methods). Each NV has an $S$=1 electron spin with a zero-field quantization axis ($\hat{S}_\mathrm{NV}$) that is along the diamond (111) axis, angled by $\alpha = 54.7$ degrees relative to the sample-plane normal (Fig.~1a). The Zeeman splitting of the NV spin states $m_\mathrm{s}=\pm1$ is induced by a small external field ($B_\mathrm{ex}\sim$ 5.6~mT) and is modulated by the projection of the local magnetic stray field ($B_\mathrm{NV}$) along $\hat{S}_\mathrm{NV}$~\cite{rondin2014magnetometry}. We detect this field by measuring the NV electron spin resonance (ESR) transitions frequencies through the application of a microwave magnetic field via a stripline and read out of the NV's spin-dependent photoluminescence (PL)~\cite{gruber1997scanning,rondin2014magnetometry}. 

As illustrated in Fig.~1b, each CrSBr layer is ferromagnetically ordered along its in-plane magnetic easy-axis (y-axis) below its N\'eel temperature~\cite{boix2022,lee2021magnetic}. Therefore, an odd number of CrSBr layers leads to a net magnetic moment that generates a magnetic stray field (purple arrows) at the CrSBr edges (Fig.~1b). We determine the projection of this field onto the NV axis (d$B_\mathrm{NV}$) by measuring the corresponding shift in the NV ESR frequency $df_\mathrm{ESR}=\gamma_\mathrm{NV}\mathrm{d}B_\mathrm{NV}$, with $\gamma_\mathrm{NV}$=28.053 GHz/T the NV gyromagnetic ratio~\cite{rondin2014magnetometry}). \textcolor{black}{From the stray field, we quantify the monolayer magnetization of the CrSBr flake and extract the NV-sample distance.} We note that the small bias field $B_\mathrm{ex}$ of 5.6~mT used in our measurements is negligible compared to the crystal anisotropy and exchange fields of CrSBr~\cite{wilson2021interlayer}, such that the CrSBr flakes preserve their interlayer antiferromagnetic ordering along the easy axis. Thus, depending on their thickness, the total stray field can have zero vs. finite value for even vs. odd numbers of the CrSBr layers.    

\begin{center}
{\large {\textbf{Diamond membrane transfer}}}
\end{center}

The main steps for the sample preparation are shown in Fig.~2 with further details in the Methods section. The diamond membrane used in this work is fabricated by ion implantation at a depth of \textcolor{black}{about} 70~nm below the diamond surface with an NV density of $10^3$ NVs/$\upmu$m$^2$ (see Methods). As shown in Fig.~2a, a large diamond chip is etched by O$_2$ plasma into squares of $50\times50~\upmu$m$^2$ with a thickness of 5~$\upmu$m, which remain connected to the surrounding diamond by small holding bars. Next, the diamond frame is attached to a metallic tip using a UV-curing adhesive. The metallic tip is held by micro-manipulators of a probe station that facilitates a controllable movement of the diamond chip. With that, the diamond chip is brought above a flexible 0.5~mm-thick polydimethylsiloxane (PDMS) layer held on a substrate. As shown in Fig.~2b, another metallic sharp tip (with $\approx$10-20~$~\upmu$m in diameter) is then used to break the connecting rod and tip out one of the square diamond membranes onto the PDMS (No.~1). \textcolor{black}{To detach the diamond membrane from the diamond frame, we first bring a target membrane to (almost) touch the PDMS surface, then push the target membrane into contact with the (sticky) PDMS polymer, then lift the frame to detach it from the membrane, and then retract the tip. This procedure eliminates the chance of the membrane flipping upside down.} The reason for choosing PDMS is to provide the diamond membrane with a rather flexible substrate with low adhesion to ease the diamond pick-up procedure that follows in the next step.  

To pick up the diamond (Fig.~2c), we use a stamp that is made of a layer of poly-methyl methacrylate (PMMA) and PDMS~(No.~2) held on a glass slide (see the methods section for the preparation of the PMMA membrane and the stamp). The PDMS-PMMA stamp is then brought in contact with the diamond membrane (held on PDMS~1), using a transfer stage that is equipped with micromanipulators. The good adhesion of the diamond and PMMA helps with the diamond pick-up when retracting the stamp (when in contact, the stack can additionally be annealed up to 80~$^\circ$C to promote the adhesion). In the next step (Fig.~2d), the diamond membrane is transferred on top of a CrSBr flake previously placed next to the Au stripline (see Methods for details).

When aligned, the diamond membrane and CrSBr flake are brought in contact while heating up the stage to 100~$^\circ$C. After cooling down the stage to 30~$^\circ$C and retracting the PDMS, the PMMA and diamond get locally detached from the PDMS in the central area and stay on the SiO$_2$ substrate. At this step, the stage is heated up to 180~$^\circ$C to melt the PMMA which relaxes the PMMA-diamond on the CrSBr-substrate and allows for full detachment of PMMA from the PDMS stamp (Fig.~2e). The PMMA in the region of interest is then removed by e-beam lithography, leaving an overlap between the unexposed part of the PMMA and the diamond membrane to avoid diamond displacement in the PMMA-removal procedure. \textcolor{black}{The PMMA removal is not necessary and is done to avoid the possibility of light scattering. We note that the presence of the PMMA has minimal effect on the detected NV response (see SI, section S5). Thus, the last step could be omitted, especially to avoid air exposure in the case of air-sensitive magnetic materials.} Note that the dark shadow around the central area of the diamond (in Fig.~2e) is a signature of the diamond membrane being directly in touch with the CrSBr flake~\cite{bogdanovic2017robust} which is essential for detecting the magnetization. The fringes observed in the optical image are related to the slight gradual variation of the diamond thickness due to the deep etching process of the diamond top surface (see SI, section S1) which does not change the NV-CrSBr distance. 

The direct contact with the intact interface between the diamond and the 2D magnet is one of the main advantages of this technique. The method minimizes the chance of contamination (e.g. dust particles) at the interface that has been affecting measurements in previous reports with a wet transfer of the diamond~\cite{bertelli2020magnetic}. Moreover, this diamond transfer method can be used in the inert atmosphere of a glove box which makes it suitable for air-sensitive 2D magnetic materials. In addition, with the full coverage of the 2D flakes with the diamond \textcolor{black}{and PMMA} membranes, further encapsulation of air-sensitive flakes can be avoided.

\begin{center}
{\large {\textbf{NV electron spin resonance measurements}}}
\end{center}

The CrSBr flake that we study here consists of regions with various thicknesses that can be distinguished by their optical contrast with respect to the $\mathrm{SiO_2}$/Si substrate (Fig.~3a). The atomic force microscope (AFM) image of the flake (Fig.~3b) and the corresponding AFM height profile (Fig.~3c) show the thickness of different regions of the CrSBr flake across its width. Over the same region, we measure the NV ESR transition frequency (Fig.~3d).  The spatial alignment of Fig.~3a-d along the width of the CrSBr flake guides us to determine the ESR signal corresponding to each step in the CrSBr flake. The non-zero modulation of the ESR frequency at some of the edges is a signature for an odd number of CrSBr flakes that would give rise to an uncompensated stray field. For such measurements, as described earlier, we are able to selectively detect the PL response from the NV centers and the CrSBr flake because of their distinct PL spectrum (Fig.~3e and SI section S1).

In Fig.~3f, we show the spatial map of the PL and ESR contrast measured at 70~K for the regions shown in the AFM image in this panel. The PL intensity map indicates a noticeable contrast as the CrSBr thickness changes considerably. On the contrary, the measured ESR signal shows insignificant variation at the large step height of CrSBr ($\approx$~28 and 72~nm) and changes considerably across the other edges with the thickness of $\sim$\textcolor{black}{2.4} and $\sim$15~nm. This is also confirmed by the same magnitude of the ESR signal measured at another \textcolor{black}{2.4 nm} CrSBr step in the middle of the flake (see panels~c,d, and g) \textcolor{black}{which corresponds to a tri-layer CrSBr edge~\cite{goser1990magnetic,lee2021magnetic}} . Moreover, our AFM characterization indicates two more steps in the CrSBr flake with \textcolor{black}{$1.1\pm0.23\,$nm and 1.71$\pm0.19\,$nm} thicknesses \textcolor{black}{associated with monolayer and bilayer of CrSBr} (shown in Fig.~S2 in SI). We observe a finite modulation of the ESR signal at the \textcolor{black}{monolayer} CrSBr step and no ESR modulation at the \textcolor{black}{bilayer} CrSBr. We conclude that the ESR measurement is indeed detecting only the uncompensated stray field generated by the steps in the CrSBr thickness corresponding to an odd number of CrSBr layers, as expected for an antiferromagnetic interlayer stacking order \textcolor{black}{\cite{healey2022varied}}. The same magnitude of the ESR modulations at different steps is consistent with the detected stray field being generated by an uncompensated magnetic moment of a single layer of the CrSBr. 

At the positions 35 $\upmu$m and ~42 $\upmu$m (Fig.~3d), we observe a modulation of $f_\text{ESR}$ of approximately double the amplitude as the modulations observed at the other steps discussed above. These positions are located in a region where the AFM image shows strong topography changes. In particular, the higher-contrast AFM image in Fig.~S2 (in SI) shows that the CrSBr in the region between ~36-45 $\upmu$m is broken into multiple smaller pieces. If two such neighboring pieces have opposite magnetic order, the stray fields generated at their edges would add, which could explain the observed doubling of the ESR shift.

\textcolor{black}{To extract the magnetization, we focus on the central peak in the NV-ensemble-measured stray field of Fig.~3d, because it is well isolated from other peaks and is located in a region where the CrSBr shows a clear, small and isolated step in its height of 2.4 nm. This enables an accurate analysis of the stray field.}. By the NV ESR measurements, we extract a stray magnetic field of d$B_\mathrm{NV}\sim$ 60~$\upmu$T generated at the edge of the \textcolor{black}{2.4 nm} CrSBr step (Fig.~3h). For quantitative magnetometry, we need to know the direction of the magnetization of the 2D magnet with respect to the NV axis ($\hat{S}_\mathrm{NV}$). For the case of CrSBr, the direction of the magnetic easy axis can be readily distinguished optically, as the mechanical exfoliation of the crystals results in elongated rectangular flakes following the in-plane magnetic anisotropy axes~\cite{telford2020layered}. As shown earlier in Fig.~1b, the direction of the CrSBr magnetization and the NV axis in our sample is such that the CrSBr stray fields have a finite projection only along the z-component of the NV spins. Thus, knowing the projection of the $\vv{M}_\mathrm{CSB}$ on $\hat{S}_\mathrm{NV}$, we can extract the uncompensated saturated magnetization of CrSBr. \textcolor{black}{By fitting the stray-field peak of Fig. 3h (see SI section S3), we find $M_\mathrm{s}$=0.46(2)~T,} consistent with the $M_s=0.48(2)$ T saturation magnetization reported by SQUID measurements in a magnetic field-polarized bulk CrSBr crystal ~\cite{goser1990magnetic}.  \textcolor{black}{  In contrast to Ref.~\cite{goser1990magnetic}, the $M_\mathrm{s}$ value in our work is extracted from the stray field of an uncompensated magnetic layer on top of a multi-layer CrSBr crystal.}  \textcolor{black}{Furthermore, we extract an NV-sample distance $z_0 = 0.13(4)$ µm (see SI). This is larger than the $\sim$70 nm NV implantation depth expected from the stopping range of ions in matter (SRIM) ~\cite{goser1990magnetic}, corroborating previous work \cite{toyli2010chip} that demonstrated that the SRIM underestimates NV implantation depth.}

\begin{figure}
    \centering
    \includegraphics[width=0.48\textwidth]{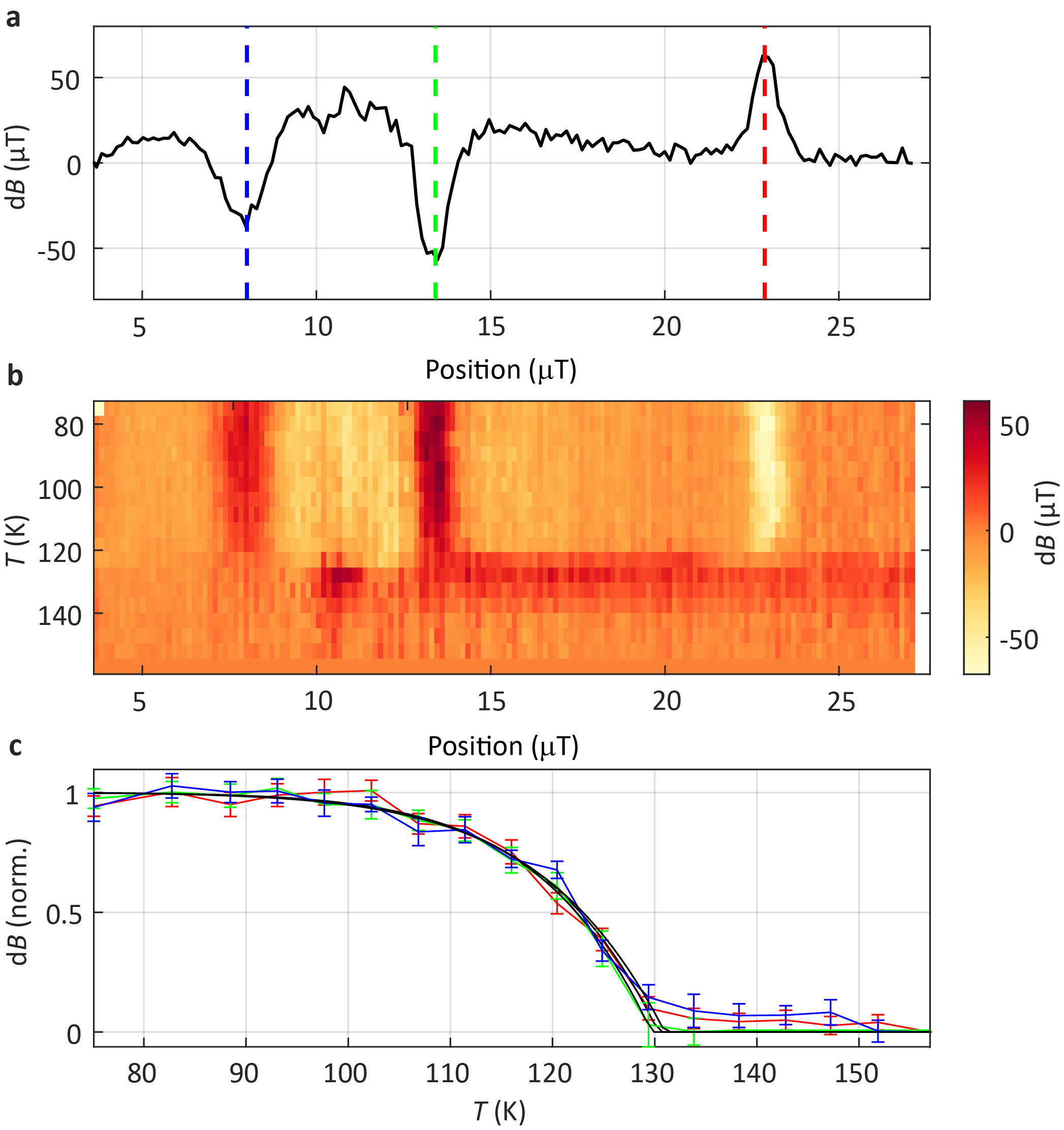}
    \caption{\textcolor{black}{Stray-field detection of the CrSBr phase transition. (a) Stray field as a function of the position below the phase transition temperature. (b) Stray field vs temperature and position showing the magnetic phase transition. For each temperature, the measured stray field at $T=157$ K was subtracted to compensate for a background variation in the ESR frequency caused by the static field inhomogeneity. (c) Normalized stray field vs temperature. By fitting with a phenomenological model $\propto 1-(T/T_c)^\beta$, we extract $\beta=11(1)$ and a phase transition temperature of $T_c=130(1)\,$K in reasonable agreement with Ref.~\cite{goser1990magnetic}.}}
    \label{fig4:my_label}
\end{figure}

\begin{center}
{\large {\textbf{Detection of magnetic phase transition}}}
\end{center}

For bulk CrSBr, the transition temperature from the antiferromagnetic to paramagnetic state is about 130~K, as measured by SQUID magnetometry~\cite{goser1990magnetic, telford2020layered}. Here we evaluate the thermal fluctuations of magnetic moments in the CrSBr flake by measuring the NV ESR frequency at various temperatures over the width of the CrSBr flake. In Fig.~4a, we show the stray field modulation across the width of the CrSBr flake measured \textcolor{black}{below the phase transition temperature} and in panel~b we show the modulation of the corresponding ESR frequency signal as a function of temperature. From there, we extract the temperature-dependence of the magnitude of the stray field at the three studied CrSBr edges, normalized by their maximum value estimated for $T$ = 0~K (d$B_\mathrm{T0}$), shown in panel~c. The d$B_\mathrm{T0}$ is estimated by the extrapolation of d$B_\mathrm{NV}$ to 0 K. As expected, the stray fields from the three CrSBr edges (with thicknesses of 15 and \textcolor{black}{2.4 nm}) show a gradual decay as the temperature is increased towards the critical temperature. We find the temperature at which the net stray fields from the edges of the CrSBr disappear to be \textcolor{black}{130 K. The extracted $T_c$ from these measurements agrees well with previously reported values~\cite{goser1990magnetic, telford2020layered}} 

\begin{center}
{\large {\textbf{Conclusions}}}
\end{center}

\textcolor{black}{The key benefits of NV magnetometry are its magnetic sensitivity, its well-understood level structure and interaction with auxiliary nuclear spins, and the precisely tunable NV density and implantation depth. Our method integrates these advantages with 2D materials assembly methods.} The developed method for the pick-up/transfer of NV-diamond membranes allows for high precision in alignment and direct contact of the diamond and magnetic target, facilitating NV-ensemble magnetometry to detect stray fields down to the monolayer limit. The compatibility of this method with the conventionally used transfer stages and glove-box inert atmosphere will pave the way for further easy and fast exploration of sensitive 2D magnetic materials. Using this technique we have obtained an intact interface at the 2D magnet-diamond interface, resulting in a spatially resolved magnetization profile. From these first-time ODMR measurements on CrSBr, we have quantified the magnetization of a single layer of the CrSBr without the need for exfoliation of a monolayer of the crystal or applying a large external magnetic field. These measurements performed on various CrSBr thicknesses resolve the uncompensated magnetization of the crystal, determining an even versus an odd number of CrSBr layers. Moreover, the temperature dependence of the measured ESR signal suggests that the stray fields generated at the edges of the 2D magnets can be more sensitive to thermal fluctuations. These experimental realizations enabled by the diamond transfer technique open the way for more applications of NV-ensemble magnetometry of 2D magnetic materials, enabling large-area and fast readout of magnetic configurations in 2D spintronic devices.\\

\begin{center}
{\large {\textbf{Methods}}}
\end{center}

\textit{Diamond Membrane Preparation.} We fabricate the $50 \times 50 \times 5$ $\upmu$m$^3$ diamond membrane (shown in Fig. 2a and 2e) from a $4 \times 4 \times 0.5$~mm$^3$ electronic-grade diamond chip acquired from Element 6 Inc. We start by having the chip cut and polished into 2 mm $\times$ 2mm $\times$ 50 $\upmu$m membranes by Almax Easylabs. After cleaning in nitric acid, we have nitrogen ions implanted by Innovion at a density of $10^{13}$/cm$^2$ and an energy of 54~keV, yielding an estimated implantation depth of 70$\pm$10~nm~\cite{pezzagna2010creation}. To create NVs, we vacuum anneal the diamond at a pressure of ~3$\times10^{-6}$ mBar by a 6-hour ramp to 400 deg., 4-hour hold, 6-hour ramp to 800 deg, 2-hour hold and cool down to room temperature. Assuming a nitrogen-to-NV conversion efficiency of ~1\% \cite{pezzagna2010creation} during the anneal, we estimate the resulting NV density to be ~$10^3$ NVs/$\upmu$m$^2$.  

We etch the microsquares into the NV-side of the diamond using a Ti mask and reactive ion etching (RIE). We deposit 50 nm of Ti using e-beam evaporation, define its shape in a PMMA resist using e-beam lithography, and etch it into shape by 3 min. of reactive ion etching in a SF$_6$/He plasma at $P_\mathrm{RF}=30 W$ RF power. We then transfer the Ti mask 5 $\upmu$m into the diamond using RIE in an O$_2$ plasma ($P_\mathrm{RF}=90$ W, $P_\mathrm{ICP}= 1100$ W, etch rate 0.25 $\upmu$m/min). As the final step, we flip the diamond and etch from the backside using O$_2$ RIE and a quartz hard mask with a square 1.4 mm opening until microsquares are released and only remain attached to the bulk diamond via a small holding bar. The Ti is removed using HF \textcolor{black}{\cite{challier2018advanced}}.

\textit{PMMA-PDMS stamp Preparation.} For the preparation/suspension of the PMMA membrane, first, we spin-coat a layer of a water-soluble polymer, Electra 92 (AR-PC 5090, All-resist) at the rate of 1000 rpm for 3 min. The Electra is then backed on a hotplate at 100~$^{\circ}$C for 1 min. Then two layers of PMMA (4$\%$, 950K) are successively spin-coated on top of the Electra layer with 1000~rpm for 3~min. After each spin-coating step, the PMMA layer is annealed on the hot plate at 180~$^{\circ}$C for 45~s. At this stage, the layer PMMA ($\approx 0.7-1 \upmu$m thick) is solidified and is ready for suspension. For that, we use a Scotch tape and cut an open window (7x7~mm$^2$) in its center. We attach the tape to the PMMA-Electra-SiO$_2$ sample and immerse the sample in water. The tape holds the PMMA membrane and gives some control over the movement of the sample, while the Electra layer gets dissolved in water. By that, the SiO$_2$ sample gets detached and the PMMA layer stays on the water surface. In the next step, we take the tape and the PMMA out of the water beaker, let it dry in air for about 10 min, and put it on top of a PDMS layer held on a grass slide~\cite{kaverzin2022spin}.

\textit{Sample Preparation.} Before using the described technique for the transfer of the diamond membrane on the CrSBr flake, there are a few steps taken for the sample preparation. The Au-stripline is made by e-beam lithography and deposition of the Ti (5~nm) and Au (100~nm) using e-beam evaporation at ultra-high vacuum ($<10^{-6}$~mbar), followed by liftoff in acetone and O$_2$ plasma treatment to remove the PMMA layer and its residues, respectively. The CrSBr flakes are cleaved from their bulk crystals using Nitto tape and are directly exfoliated on a PDMS layer. Then they are transferred from the PDMS on the target substrate distanced by $<5~\upmu$m from the Au stripline using micro-manipulators of a transfer stage. The sample preparation could also be done by the lithography of the Au stripline directly on the substrate where the CrSBr flake is exfoliated on. 

\textit{CrSBr synthesis.} CrSBr crystals are prepared by the direct reaction of their components in a stoichiometric ratio, mixing chromium (99.99 \%, Alfa-Aesar), sulfur (99.99 \%, Sigma-Aldrich), and bromine (99.9 \%, Sigma-Aldrich), with a 3 \% in mass excess of bromine, which acts as a transport agent. Crystals are characterized by powder and crystal X-Ray diffraction, energy dispersive X-Ray analysis (EDX), high-resolution TEM, SQUID magnetometry and temperature-dependent single crystal, as reported in Ref.~\cite{boix2022}.

\textit{Photoluminescence microscope setup.}
We detect the NV photoluminescence (PL) using a home-built cryogenic microscope setup. The cryostat is a Montana S100 hosting a room-temperature-stabilized NA = 0.85 microscope objective. We position and focus the sample using cryogenic slip-stick positioners (ANPx101 and ANPz101, Attocube), and use a fast steering mirror (FSM300, Newport) to make spatial NV PL maps (Fig.~3d, Fig.~3f). We excite the PL using a continuous-wave 520 nm laser (Obis LX 520, Coherent) and detect it using an avalanche photodiode (Excelitas, SPCM-AQRH-13) after suppressing the laser and the CrSBr PL by band-pass filtering (600-800 nm). The NV+CrSBr PL spectrum shown in Fig. 3e is recorded with an Andor Kimera spectrometer equipped with an Ivac CCD camera using a 550 nm long-pass filter. A Windfreak SynthHDv2 generates the  microwaves used for driving the NV spins.

\begin{flushleft}
{\large {\textbf{Acknowledgements}}}
\end{flushleft}

We would like to acknowledge the assistance of Allard J. Katan with this project. This project has received funding from the European Union Horizon 2020 research and innovation program under grant agreement No. 863098 (SPRING), and from the Netherlands Research Council (NWO) under grants 740.018.012 (startup), 680.91.115 (Projectruimte) and 193.077 (VIDI). SMV and CBC acknowledge  the financial support from the European Union (ERC AdG Mol-2D 788222) and the Spanish MCIN (2D-HETEROS PID2020-117152RB-100, co-financed by FEDER, and Excellence Unit “María de Maeztu” CEX2019-000919-M). SMV thanks the Generalitat Valenciana for a postdoctoral fellow APOSTD-CIAPOS2021/215.   

\begin{flushleft}
{\large {\textbf{Author Contribution}}}
\end{flushleft}

T.S.G., S.K., H.S.J.v.d.Z., and T.v.d.S. conceived and designed the experiments. M.B. and B.G.S. fabricated the diamond membranes. S.K. fabricated the stripline and developed the diamond tipping process. T.S.G. developed the diamond membrane transfer process, performed AFM characterizations, and fabricated the CrSBr-diamond sample. C.B. and S.M. synthesized the CrSBr crystals. M.B. performed the ESR measurements, with the help of S.K., and I.B.. T.v.d.S. analyzed the ESR data. T.S.G. wrote the manuscript with contributions from all co-authors.\\

\renewcommand{\figurename}{Fig.}
\renewcommand{\thefigure}{S\arabic{figure}}
\setcounter{figure}{0}

\newpage
\textcolor{white}{Hi}
\newpage

\begin{widetext}

{\Large \textbf{Supporting Information:}} \\
{\large \textbf{Nitrogen-vacancy magnetometry of CrSBr by diamond membrane transfer}}\\
\\

{\large \textbf{S1.~Optical micrographs and photoluminescence measurements}}\\

\begin{figure*}[h!]
    \centering
    \includegraphics[width=0.8\textwidth]{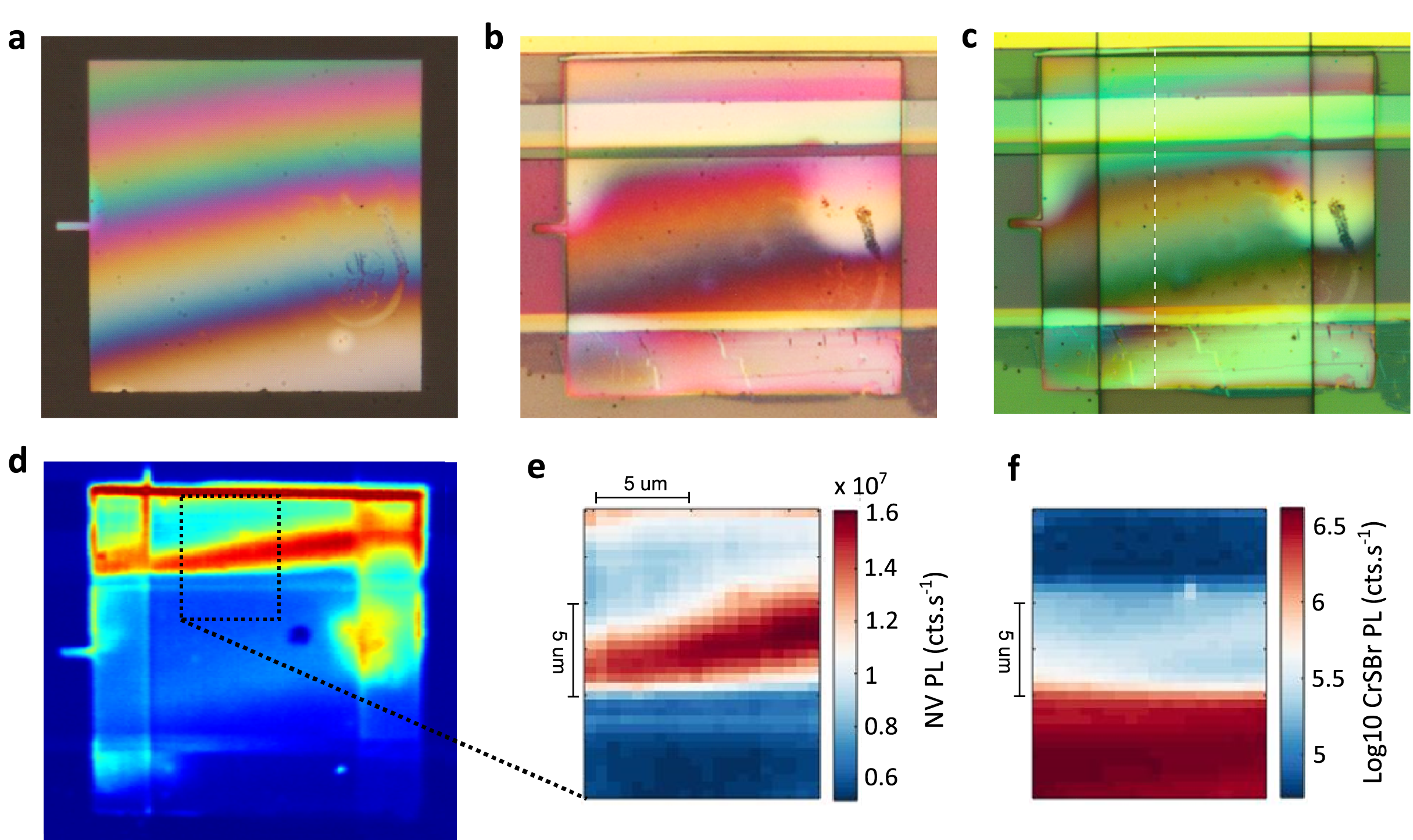}
    \caption{(a)~Optical micrograph of the diamond membrane on PDMS. The fringes observed on the diamond membrane are due to the gradual change in the diamond thickness because of the deep etching process of the top surface. Note that this thickness variation does not affect the 70~nm distance between the NV centers and the other surface of the diamond membrane (that comes in contact with the CrSBr flake). (b)~Optical micrograph of the diamond membrane transferred on top of the CrSBr flake, covered with the PMMA layer used for the transfer. (c)~Optical micrograph of the diamond membrane-CrSBr after selective removal of the PMMA layer by e-beam lithography. \textcolor{black}{The white dashed line indicates the location of the measurements shown in the main manuscript, Fig.~3d and Fig.~4a}. (d) spatially-resolved photoluminescence (PL) response of the diamond membrane. (e,f) PL response of the NV centers and CrSBr flake for the region shown by the black dashed square. \textcolor{black}{Panel d shows a sudden increase of the NV PL near the top of the image. We believe this is caused by the sudden separation between the diamond and the CSB, which in turn is due to a step in the CSB thickness (see the 120 nm-high plateau in the AFM linetrace of Fig. 3c in the main manuscript). Such separation reduces the outcoupling of the NV photoluminescence towards the sample side because of the lower refractive index of vacuum, and correspondingly increases the NV photoluminescence radiated towards our collection optics.}  }
    \label{fig:myPL}
\end{figure*}

\newpage
{\large \textbf{S2.~Atomic force microscopy (AFM) measurements }}\\

\begin{figure*}[h!]
    \centering
    \includegraphics[width=0.65\textwidth]{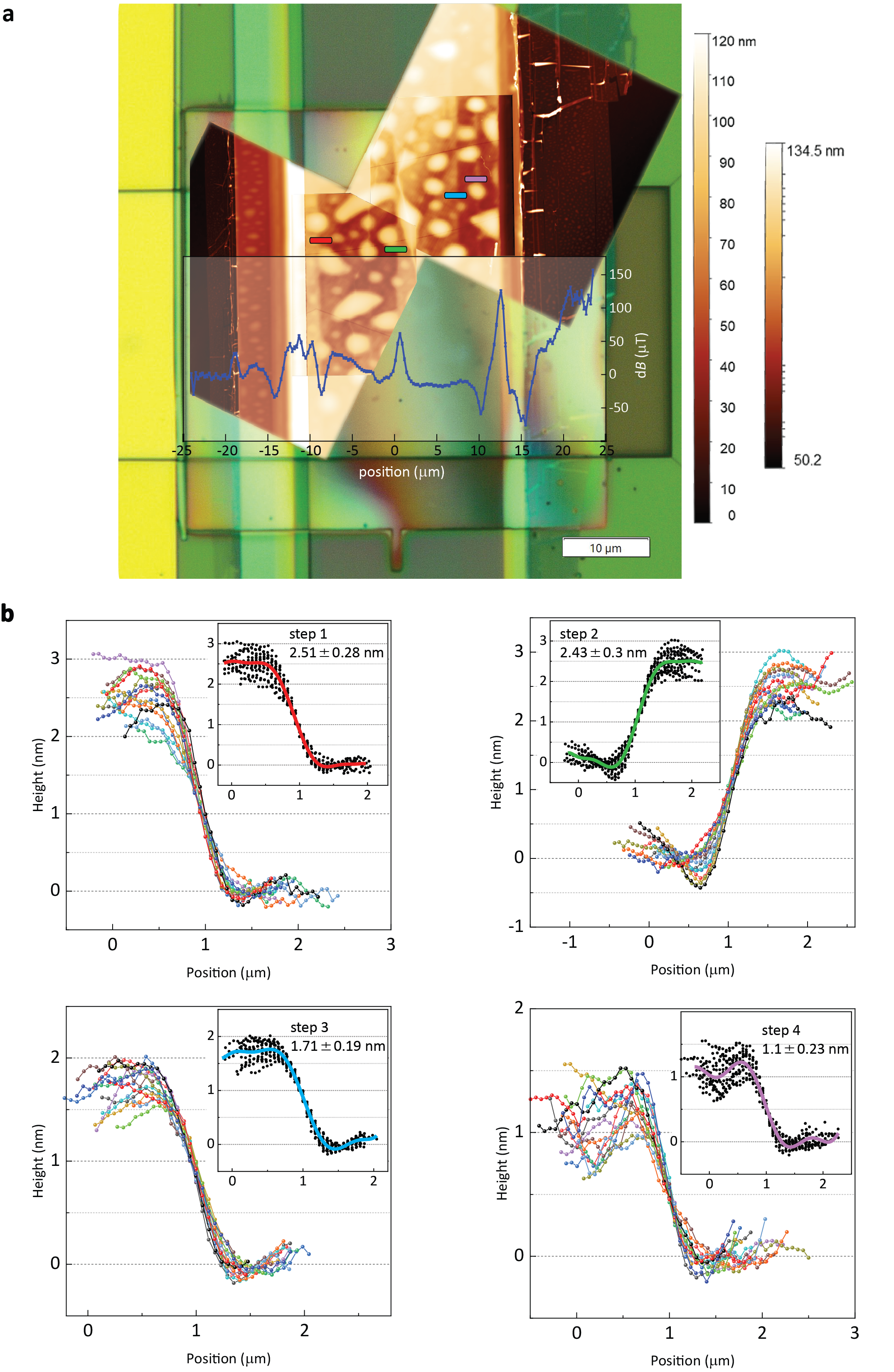}
    \caption{(a) AFM images of the studied CrSBr flake overlapped with the optical image of the sample (after diamond transfer). The larger AFM images are shown in linear scale and the smaller AFM images are presented in an adaptive nonlinear color range to highlight the few-nanometer-thin features in the CrSBr flake. The blue plot is the d$B_\mathrm{NV}$ extracted from the ESR measurement along the width of the CrSBr channel, overlapped with respect to the other images to highlight the exact locations on the CrSBr flake that give rise to the modulations of the detected stray fields. (b) AFM height profiles measured for the \textcolor{black}{mono-, bi- and tri-layer} CrSBr steps along the lines in panel a, colored correspondingly. \textcolor{black}{For each step, we extract the height profiles taken from different regions of the flake along the whole edge of the step, excluding the bubbles. For each step, we take 15 traces and offset each so that they start from the same background level. The height value reported in the plots is extracted by taking the mean height value of all 15 traces and subtracting the mean value at the base height. The error bars show the standard deviation from the mean values. }} 
    \label{fig:myAFM}
\end{figure*}

In Fig.~\ref{fig:myAFM}, we show the overlap of the AFM height profile images and the optical image of the sample. By presenting the height measurements in the AFM images by adaptive nonlinear color mapping we highlight the steps in the CrSBr thickness (confirmed by the height profile of the edges in panel~b). The clear correspondence of the modulations of the detected stray field (d$B_\mathrm{NV}$) with respect to the location of the edges in the AFM and optical images, show the finite d$B_\mathrm{NV}$ detected for the \textcolor{black}{mono-layer and tri-layer} CrSBr edges and its absence for the \textcolor{black}{bi-layer CrSBr} edge. 

We note that the AFM image of the CrSBr flake (Fig.~\ref{fig:myAFM}) shows the presence of bubbles at the interface between the flake and SiO$_2$ substrate. To assess a potential effect on the NV-sample distance, we characterized the dimensions of 20 bubbles in the AFM image. We extract a mean bubble height of 10$\pm$4~nm and a mean lateral bubble diameter of 3$\pm$1~$\upmu$m, i.e., more than two orders of magnitude larger than the bubble height. This extreme flatness of the bubbles presumably allows for a slight conformation of the diamond membrane such that it locally stays in contact with the CrSBr. The absence of the spatial features in our PL and ESR measurements (Fig.~3, main manuscript) further indicates that the NV-CrSBr distance is not affected by the presence of bubbles.     \\

{\large \textbf{S3.~Extracting Ms from the ensemble ODMR frequencies}}\\

\begin{figure*}[h!]
    \centering
    \includegraphics[width=\textwidth]{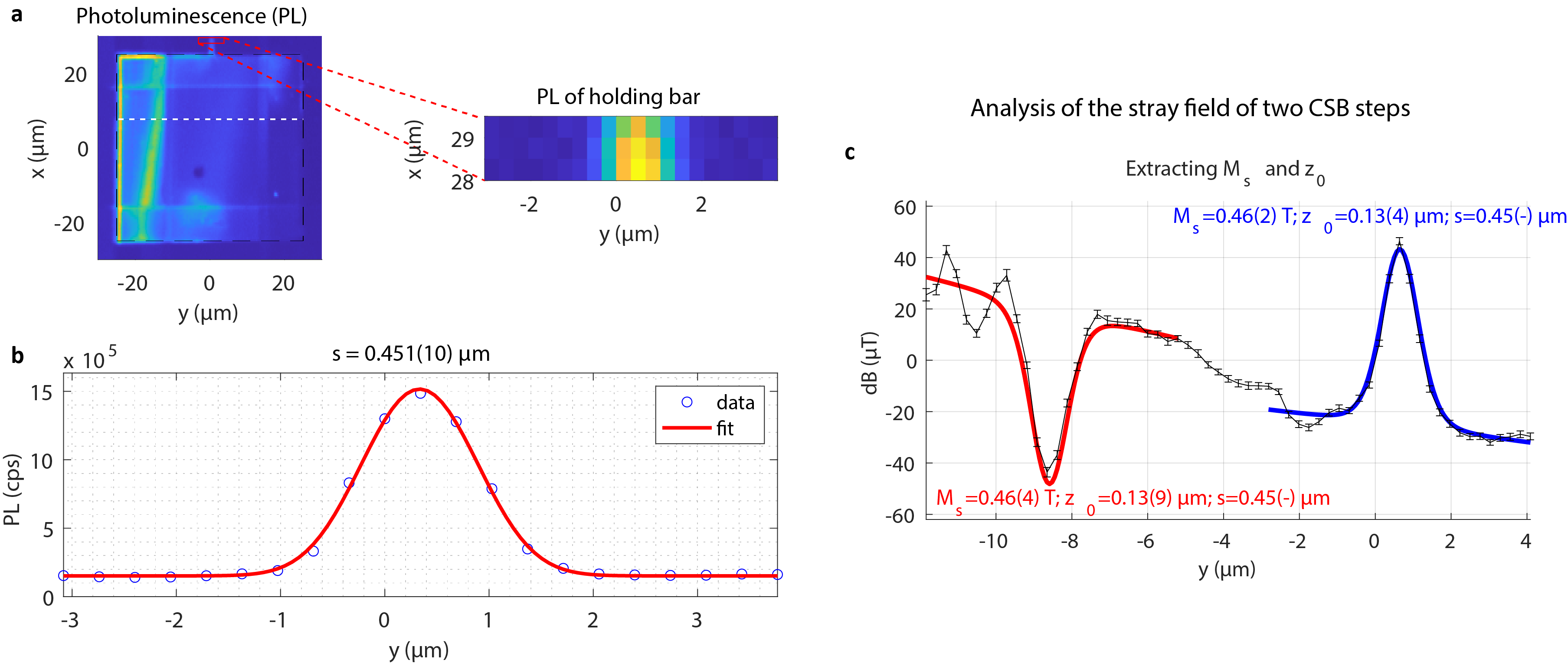}
    \caption{\textcolor{black}{\textbf{Analysis of the NV-ensemble measured stray field $dB$ at two magnetization steps. }\textbf{(a)} Photoluminescence (PL) map of the NV-diamond on the sample. The zoom-in shows the PL of the 1 $\mu$m-wide holding bar from which we extract the optical spot size $s$. The white dashed line indicates the location of the measurements shown in (c). \textbf{(b)} Measured and fitted PL across the holding bar. The fit was performed by convolving a Gaussian function $e^{-x^2/2s^2}$ with the 1-$\mu$m holding-bar width, yielding $s=451(10)$ nm. \textbf{(c)}. Fit of the stray field above two CSB steps from which we extract the CSB magnetization $M_s$ and the NV-sample distance $z_0$, with $s=0.45$ µm fixed at its independently measured value (see (b)). The fitted values are shown in the plot. The well-isolated peak on the right is the same as the one in Fig. 3h of the main text.}}
    \label{fig:FigR1}
\end{figure*}

As described in the main text, we extract the CrSBr magnetization $M_\mathrm{s}$ by analyzing the magnetic stray field at the \textcolor{black}{2.4~nm} step in the CrSBr flake indicated in Fig.~3h of the main manuscript. Here we describe the fitting procedure. The step is oriented parallel to the $x$-axis and located at $y=y_0$. For the small magnetic bias fields used, the CrSBr magnetization is expected to be oriented along $y$ (easy axis), i.e., perpendicularly to the step edge. If the step corresponds to a change in the CrSBr thickness by an odd number of layers, the uncompensated layer magnetization leads to a magnetic stray field that emerges from the edge. The $z$ component of this field in the NV layer is given by
\begin{equation}
B_\mathrm{z}(y) = \frac{ M_\mathrm{s} t}{2 \pi}\frac{z_0}{z_0^2+(y-y_0)^2}
\end{equation}
where $t=0.79$ nm is the thickness of a single CrSBr layer~\cite{lee2021magnetic}, $M_\mathrm{s}$ is the magnetization in Tesla, and we used $t \ll z_0$ with $z_0$ the NV-implantation depth.\\

The projection of $B_\mathrm{z}(y)$ on the NV-axis $dB_\text{NV}(y) = B_\mathrm{z}(y)\cos(54.7^\circ)$ leads to a shift in the ESR frequency of the NV spins given by $df_\text{ESR}(y) = \gamma dB_\text{NV}(y) $. We characterize this shift by measuring the optically detected magnetic resonance (ODMR) spectra of our NVs. We do so by sweeping the frequency of a microwave drive applied to the stripline and detecting the spin-dependent NV photoluminescence. The normalized ODMR response of a single NV spin at location $y$ is described by the Lorentzian 
\begin{equation}
S_\mathrm{s}(y, df) = 1- \frac{1}{(df-df_\text{ESR}(y))^2 + w^2}
\end{equation}
where $df$ is the detuning between the microwave drive and the `offset' ESR frequency associated with the applied bias field, and $w$ is the width of the ESR response. The measured ensemble ODMR response is smeared by our diffraction-limited optical spot size, as described by the convolution of $S_\mathrm{s}(y)$ with a Gaussian:
\begin{equation}
S(y, df) = S_\mathrm{s}(y, df)  * \frac{1}{\sqrt{2\pi s^2}} e^{-y^2/2 s^2}
\end{equation}
with $s$ the standard deviation of the optical spot. We extract $df_\text{ESR}(y)$ by fitting $S(y,df)$ at each $y$-location with a Lorentzian. We then fit these values for $df_\text{ESR}(y)$ to the experimentally extracted values, which we obtain by fitting lorentzians to the measured ensemble ODMR curves.

\textcolor{black}{To extract the NV-sample distance $z_0$ from the stray-field data, we need to know the optical spot size $s$. Therefore, we first determine $s$ by the analysis shown in Fig.~\ref{fig:FigR1}. We do so by fitting the photoluminescence across the diamond holding bar (Fig.~\ref{fig:FigR1}a) with a Gaussian $e^{-x^2/2s^2}$ convolved with the 1-$\mu$m-wide holding-bar width, yielding $s=451(10)$ nm (Fig.~\ref{fig:FigR1}b). We then fix $s$ to this value and extract both $M_s$ and $z_0$ by fitting the stray field (Fig.~\ref{fig:FigR1}c). These fits yield $M_s=0.46(2)$ T and $M_s=0.46(4)$ T. Furthermore, we extract $z_0 = 0.13(4)$ µm and $z_0=0.13(9)$ µm for the two peaks - which is indeed larger than the $\sim$70 nm expected from SRIM.}

\newpage

{\large \textbf{\textcolor{black}{S4.~Stray-field detection of the CSB phase transition at different microwave and laser powers}}}\\
\begin{figure*}[h!]
    \centering
    \includegraphics[width=0.8\textwidth]{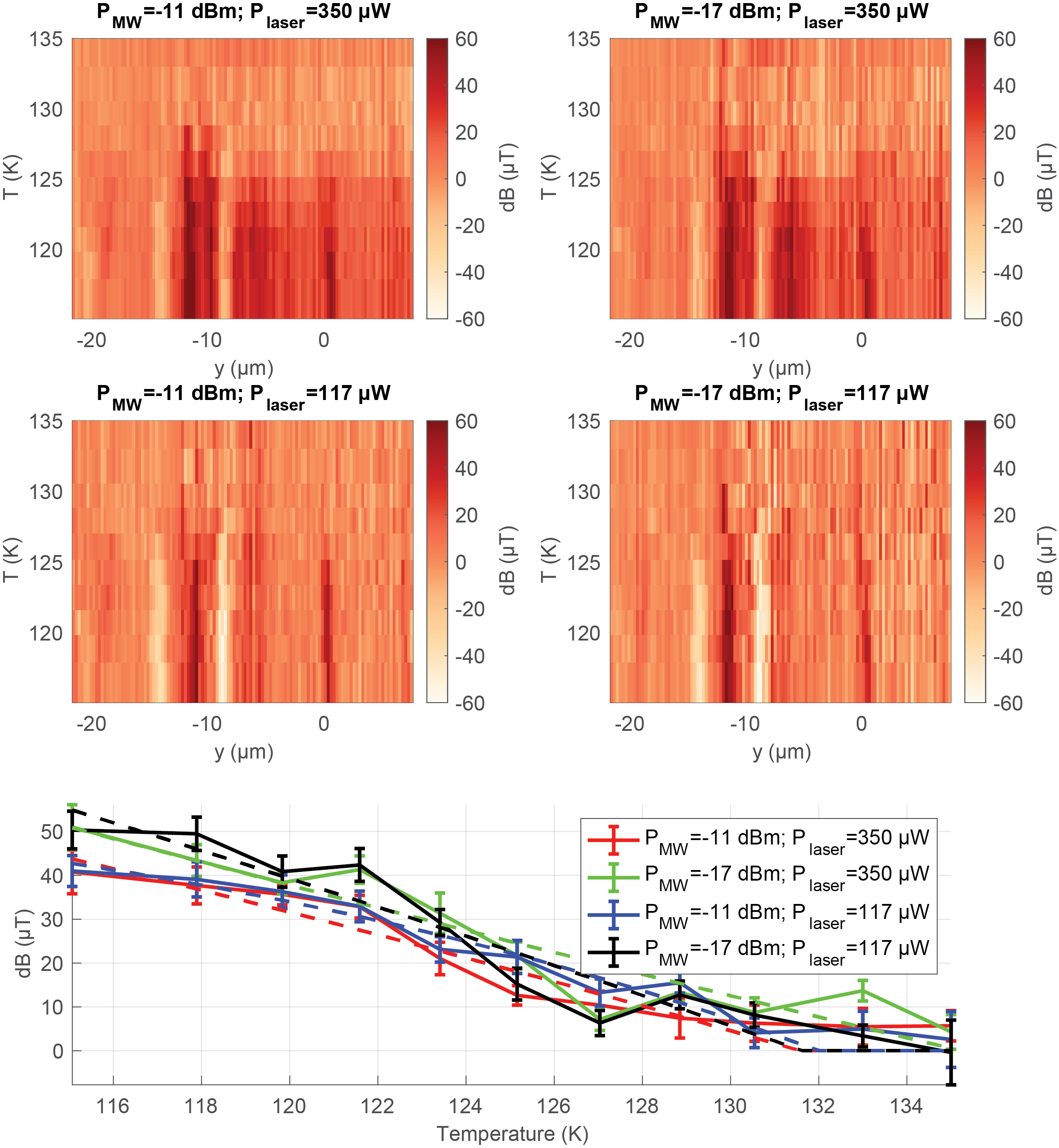}
    \caption{\textcolor{black}{\textbf{Stray-field detection of the CSB phase transition at different microwave and laser powers.} (a) Stray field as a function of position and temperature at different laser and microwave powers. (b) Amplitude of the stray-field peak at $y=-14.5$ µm as a function of temperature at different laser and microwave powers. There is no significant dependence of the stray-field amplitude on the laser and microwave powers.} }
    \label{Fig:power-dep}
\end{figure*}

\newpage

{\large \textbf{\textcolor{black}{S5.~Effect of PMMA-coverage on the NV-related signal}}}\\
\begin{figure*}[h!]
    \centering
    \includegraphics[width=1\textwidth]{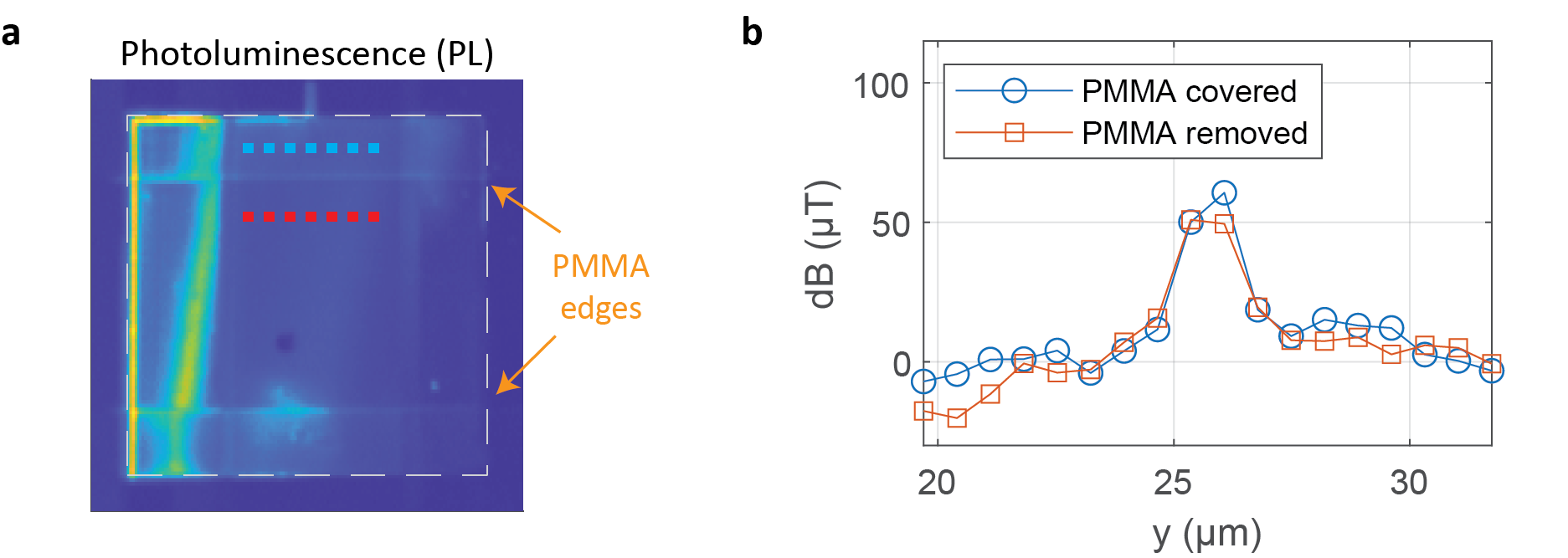}
    \caption{\textcolor{black}{(a) Spatially-resolved PL response of the diamond membrane, with the blue (red) dashed lines indicating the position with (without) the PMMA membrane, where the ESR frequency of panel b is measured. (b) Comparison of the amplitude of the ESR signal detected at the 2.4 nm step in the regions that are with and without the PMMA membrane}}
    \label{Fig:PMMA}
\end{figure*}

\end{widetext}

\end{document}